\documentclass{IAU-TrA}
\usepackage{graphicx}\usepackage{color}

\title[STELLAR CONSTITUTION]     
{}

\author[DIVISION~IV / COMMISSION 35]   
{}

\pubyear{2012}
\volume{Volume XXVIIIA}
\pagerange{001--008}
\jname{Transactions IAU, Volume XXVIIIA}
\editors{Ian Corbett, ed.}
\begin{document}

\maketitle

{\bf

\large
\begin{tabbing}
\hspace*{65mm}       \=                                              \kill
COMMISSION~35         \> STELLAR CONSTITUTION                            \\
                     \> {\small\it CONSTITUTION DES ETOILES}                 \\
\end{tabbing}

\normalsize

\begin{tabbing}
\hspace*{65mm}       \=                                              \kill
PRESIDENT            \> Corinne Charbonnel                                \\
VICE-PRESIDENT       \> Marco Limongi                             \\
PAST PRESIDENT       \> Francesca D'Antona                         \\
ORGANIZING COMMITTEE \> Gilles Fontaine, Jordi Isern,              \\ 
                     \> John C. Lattanzio, Claus Leitherer,          \\
                     \> Jacco Th. van Loon, Achim Weiss,           \\
                     \> Lev R. Youngelson                                      \\   
\end{tabbing}

\noindent
COMMISSION~35 WORKING GROUPS
\smallskip

\begin{tabbing}
\hspace*{65mm}      \=                                               \kill
Div.~IV / Commission~35 WG \>  Active B Stars                  \\
Div.~IV / Commission~35 WG \>  Massive Stars                       \\
Div.~IV / Commission~35 WG \>  Red Giant Abundances                          \\
Div.~IV / Commission~35 WG \>  Chemically Peculiar and Related Stars                               \\
\end{tabbing}

\bigskip

\noindent
TRIENNIAL REPORT 2009-2011
}

\firstsection 

\section{The activity}

Commission 35 consists of members of the International Astronomical Union whose research is concerned with the structure and evolution of stars in all parts of the H-R diagram. Their interests range from various aspects of stellar interior physics, such as convection, diffusion, rotation, magnetic fields, to asteroseismology and the prediction of the evolutionary and nucleosynthetic histories of stars that are of vital importance for our understanding of stellar populations and galactic chemical evolution.

After discussion with the OC and with many members of C35, the present President, Corinne Charbonnel, proposes to change the name of C35 to ``Stellar structure and evolution". We shall ask reactions to C35 members in spring 2012, and we expect that this proposition will be accepted by IAU during the next GA in Beijing.

Commission 35 is part of Division IV of the International Union (Stars), which is concerned with the characterization, interior and atmospheric structure, and evolution of stars of all masses, ages, and chemical compositions.   
Division IV acts as the umbrella for Commissions 26 - Double and Multiple Stars, 29 - Stellar Spectra, 35 - Stellar Constitution, 36 - Theory of Stellar Atmospheres, and 45 - Stellar Classification. 
Division IV has established four Working Groups: WG on Active B Stars, WG on Massive Stars, WG on Red Giant Abundances, and WG on Chemically Peculiar and Related Stars.

The Commission home page (http://iau-c35.stsci.edu) is maintained by Claus Leitherer and contains general information on the Commission structure and activities, including links to stellar structure ressources that are made available by the owners. The  The resources contain evolutionary tracks and isochrones from various groups, nuclear reactions, EOS, and opacity data, as well as links to main astronomical journals.

Members of IAU Commission 35 are encouraged to take advantage of a discussion board for communicating with their colleagues and friends. This service is moderated and maintained by the Organizing Committee of Commission 35 and hosted by http://iau-c35.stsci.edu/Blog/index.html.

As a routine activity, the Organizing Committee has commented on and ranked proposals
for several IAU sponsored meetings. 
Our Commission  participated in the organization of the following Symposia, Joint Discussions, and Special Sessions held at the XXVIth IAU General Assembly in Rio in August 2009: IAUS 262 ``Stellar Populations - Planning for the Next Decade", JD3 ``Neutron Stars: Timing in Extreme Environments", JD4  ``Progress in Understanding the Physics of Ap and Related Stars", SpS7 ``Young Stars, Brown Dwarfs, and Protoplanetary Disks", JD11``New Advances in Helio- and Astero-Seismology", JD13 ``Eta Carinae in the Context of the Most Massive Stars", SpS1 ``IR and sub-mm spectroscopy: a new tool for studying stellar evolution".

Members of Commission 35 were involved in organization of other IAU sponsored
meetings: IAUS 268 ``Light Elements in the Universe", IAUS 271``Astrophysical dynamics Ð from stars to galaxies", IAUS 272 ``Active OB stars Ð structure, evolution, mass loss, and critical limits", IAUS 281``Binary Paths to the Explosions of type Ia Supernovae", IAUS 283 ``Planetary Nebulae: an Eye to the Future".
Many other international meetings, in which members of the Commission were involved, were held in these years.

In the following we present some highlights that were published during the present triennial term and which concern the field of ``Stellar Constitution". 

\section{Transport processes in stars (C. Charbonnel)}
\label{transportCC}

It is now widely recognised that the art of modelling stars strongly depends on our hability to model the various processes that transport angular momentum and chemicals in stellar interiors. This is a prerequisite to reproduce detailed data in various parts of the HR diagram, especially in the new area where asteroseismology revolutionises our research field (see \S\ref{sect:astero}). In the recent years important advances were made to take into account into stellar models the physics of atomic diffusion, rotation, magnetic fields, internal gravity waves, mass loss, and of various hydrodynamical instabilities among which thermohaline convection. This is a challenging task since such mechanisms involve length and time-scales that differ by several orders of magnitude and impact on the stars both on dynamical and secular times. It implies that one needs to use and couple 1D, 2D, and 3D approaches to get a global picture of macroscopic MHD transport processes in stellar interiors. With the great endeavour made over the past two decades, work to go beyond the classical spherical picture of stars to get a global MHD understanding of their internal dynamics is now in a golden age. From dynamical to secular time scales, 1 to 3D approaches of the macroscopic transport processes indeed allow stellar modellers to enter into the details of the highly non-linear couplings between meridional circulation, differential rotation, turbulence, fossil and dynamo magnetic fields, and waves. 
Here we present some highlights concerning magneto-hydrodynamical processes in stellar radiative regions that appeared in the literature since 2009. Other related results are described in the other sections of this report.

Rotational mixing based on the prescription by Zahn, Maeder and collaborators was shown to explain the surface abundances of Li, Be, C, N, Na in low- and intermediate-mass stars (more massive that $\sim$1.5M$_{\odot}$) belonging to open clusters and to the field (e.g., Charbonnel \& Lagarde 2010, Canto Martins et al. 2011). 
Diagnostic tools were designed to help disantangle the role of the various rotation-driven mechanisms in stars (Decressin et al. 2009), and large grids of rotating models and isochrones were computed (Brott et al. 2011; Ekstr\"{o}m et al., in press; Lagarde et al., submitted). 
The effects of rotational mixing on the asterosismic properties of the Sun (Turck-Chi\`eze et al. 2010) and of solar-type stars (Eggenberger et al. 2010) were examined in detail. The constraints on internal angular momentum transport in Solar-type stars that can be inferred from the spin-down of open cluster stars and from the solar rotation curve were critically examined; this lead to the conclusion that neither hydrodynamic mechanisms nor a revised and less efficient prescription for the Tayler-Spruit dynamo can reproduce both spin-down and the internal solar rotation profile by themselves (Denissenkov et al. 2010, Denissenkov 2010a). It is thus confirming the idea that in solar-type stars a successful model of angular momentum evolution must involve more than one mechanism. This has increased the urgency to describe the physics related to magnetic fields and internal gravity waves in stellar interiors.

Magnetic field (which are observed more and more extensively at stellar surfaces through spectropolarimetry) and their related dynamical effects are thought to be important in stellar interiors. Intense researches were devoted to understand their role in convective layers, in particular for the sun, as well as in radiative zones. Several papers were devoted to preparing the implementation in one-dimensional stellar evolution codes (i.e., the requested modifications to the structural equations of stellar evolution) of the different terms induced by the presence of a magnetic field (Duez et al. 2009, 2010), and to understand the fossil fields' origin, topology, and stability (the semi-analytical work of Duez \& Mathis 2010). This is an important step to get a true MHD approach in stellar evolution. 

Several works were devoted to study the transport of angular momentum by the internal gravity waves excited by turbulent motions at the border of stellar convective regions. 3D numerical simulations were aimed at determining the wave spectrum (Brun \& Strugarek 2010). 
As far as mode excitation by turbulent convection is concerned, the influence of the Coriolis acceleration on the stochastic excitation of oscillation modes in rotating stars was studied through a perturbative analysis (Belkacem et al. 2009b).
The transport by gravito-inertial waves has been studied in the case of a general differential rotation (Mathis 2009). 
Analytical models (MacGregor \& Rogers 2011) and numerical simulations (Rogers \& McGregor 2010, 2011) were carried out in order to investigate the interaction of internal gravity waves with magnetic fields in stellar radiative interiors. A global study of magneto-gravito-inertial waves in stellar radiation zones was achieved in the case of an axisymmetric toroidal magnetic field (Mathis \& de Brye 2011).
Also, non-linear effects such as wave-braking at the center of the star were studied; signs appear of a possible acceleration of the central rotation of solar-type stars by this mechanism (Barker \& Ogilvie 2010, 2011). In the near future, consequences for angular momentum transport and the case of general differential rotation and azimuthal magnetic field have to be studied. 

Self-consistent stellar evolution models including atomic diffusion and radiative accelerations together with other macroscopic processes were published. 
The effects of atomic diffusion on internal and surface abundances of A- and F- pre-main-sequence stars with mass loss were studied in order to determine at what age the effects materialize, as well as to further understand the processes at play in HAeBe and young ApBp stars. Atomic diffusion in the presence of weak mass loss was shown to be able to explain the observed abundance anomalies of pre-main-sequence stars, as well as the presence of binary systems with metal-rich primaries and chemically normal secondaries such as V380 Ori and HD 72106. This is in contrast to turbulence models which do not allow for abundance anomalies to develop on the pre-main-sequence. The age at which anomalies can appear depends on stellar mass (Vick et al. 2011).
A thorough study of the effects of mass loss on the internal and surface abundance of A- and F-type main sequence stars was undertaken in order to constrain mass loss rates and to elucidate some of the processes that compete with atomic diffusion in these stars (Vick et al. 2010, Michaud et al. 2011). However the current observational constraints (e.g. Gebran et al. 2010) do not allow to conclude that mass loss is to be preferred over turbulent mixing (induced e.g. by rotation) in order to explain the AmFm phenomenon, although internal concentration variations which could be detectable through asteroseismic tests should provide further information.
Last but not least, the effect of atomic diffusion was studied during the relatively rapid red giant  phase to determine the concentration variations it leads to and at what accuracy level it can be safely neglected (Michaud et al. 2010), as well as during the horizontal branch evolution (Michaud et al. 2011).

Several works were devoted to the study of the thermohaline double-diffusive instability that is expected to develop in stars under various circumstances. 
Its influence on (atomic) diffusion-induced iron accumulation was studied for A-F stars; it was shown that iron accumulation is still present when thermohaline convection is taken into account, but much reduced compared to when this physical process is neglected (Th\'eado et al. 2009, Th\'eado \& Vauclair 2010). Metal-rich accretion and thermohaline instability were investigated in exoplanets-host stars, as well as their consequences on the light elements abundances (Th\'eado \& Vauclair 2011). The combined effects of thermohaline instability and rotation-induced mixing were studied for low- and intermediate-mass stars at various metallicities; these mechanisms were shown to account for the observational constraints very well over the whole investigated mass range for stars on the main sequence, on the red giant branch and clump, and on the early-AGB (Charbonnel \& Lagarde 2010; see also Stancliffe et al. 2009, and Stancliffe 2010), and to account for new $^3$He yields that reconcile Galactic requirements for the evolution of this light elements with the Big Bang nucleosynthesis (Lagarde et al. 2011). 
In super-AGB stars thermohaline mixing becomes important after carbon has been ignited off-center and it affects significantly the propagation of the flame; however during the subsequent thermally pulsing SAGB phase, the high temperature at the base of the convective envelope prevents the development of thermohaline instabilities associated with 3He burning as found in low-mass red giant stars (Siess 2009). 
Although these evolutionary studies sound promising, the question of the efficiency of the thermohaline instability in stellar interiors remains open. In particular the geometry of the instability cells that is deduced from the linear stability analysis and that is favoured by observational data for the red giants and by laboratory experiments is that of long thin fingers with an aspect ratio of the order of 6. This value turns out to be $\sim$ 5 times higher than that obtained by current simulations of thermohaline convection (Denissenkov 2010b; Denissenkov \& Merryfield 2011; Rosenblum et al. 2011; Traxler et al. 2011). However these simulations are still quite far from the stellar regime (i.e., in the best cases they are run at moderately low values of the Prandtl number), which calls for future numerical multi-dimensional simulations. Magneto-thermohaline mixing was studied as an alternative to thermohaline convection (Denissenkov et al. 2009).
In view of the uncertainties affecting the current description of the mixing mechanisms at act in low-mass red giant stars (see also Cantiello \& Langer 2010), the question of abundance anomalies on the RGB and AGB was revisited by means of parameterized non-convective mixing in post-processing calculations (Palmerini et al. 2011a,b).  

Tidal dynamics of extended bodies in planetary systems and multiple stars was examined 
within the context of the recent discovery of a large number of extrasolar planets orbiting their parent stars at distances lower than 0.1 astronomical unit and the launch  of dedicated space missions such as CoRoT and KEPLER; this is obviously also important when considering the position of inner natural satellites around giant planets in our Solar System and the existence of very close but separated binary stars. An important step beyond the traditional approach was made by Mathis \& Le Poncin-Latiffe (2009) who derived the dynamical equations for the gravitational and tidal interactions between extended bodies and associated dynamics and gave the conditions for applying this formalism.

\section{Helio- and asteroseismology (J. Montalban \& A. Noels)}
\label{sect:astero}

Without underestimating  the contribution from ground-based observations, which are essential, for instance,  for mode identification of some stellar pulsators, the enormous  progress  recently made by asteroseismology is mainly due to the high quality light curves provided  for a huge number of targets across the HR diagram by the space satellites CoRoT (Baglin et al. 06) and {\it Kepler}  ( Borucki et al. 2009). Both missions have the dual goal of detecting extra-solar planets by the transit technique (which implies high level of photometric sensitivity) and of observing stellar oscillations.  CoRoT was launched on December 2007 and, although it was planned to operate during three years,  its excellent results have led to the extension of the mission until March 2013.  {\it Kepler} satellite was launched on March 2009 and it will operate during 3-4 years.  The features of the telescopes, orbits, and observations are different and both missions provide complementary information. In particular,  {\it Kepler} and CoRoT cover different regions of the Milky Way.  {\it Kepler} observes a fixed (and more extended) field above the Galactic plane (b=13$^o$.5) in the region of Cygnus-Lyra,  while CoRoT alternates each 5 months observation fields in the direction of the Galactic center and of the anti-center, and not too far from the Galactic plane. For each observation field both space telescopes provide several thousands of light curves. These huge numbers together with the high photometric sensitivity and long time coverage of observations are at the origin of the exceptional oscillation data and results that we summarize below.

Oscillations stochastically excited by turbulence in convective envelopes (such as those observed in the Sun -- solar-like oscillations) have been detected in several hundreds  of dwarfs and sub-giants (more than 500) and in several thousands of red giants.  The large number of targets involved in these studies has lead to what has been called ``ensemble asteroseismology"  (Chaplin 2010).

The detection of solar-like oscillations in G-K red giants has provided the most surprising results with a great  impact on other astrophysical domains. Although stochastic oscillations in red giant stars were already known from ground-base observations (i.e., Frandsen et al. 2002), their non-radial character was definitely proven by CoRoT observations (de Ridder et al. 2009), and obliged to review the previous ideas concerning solar-like oscillation in this kind of objects.  The oscillation spectra of about two thousand red giants showed a power excess with an almost gaussian shape centered on the frequency $\nu_{max}$ and a regular pattern of radial modes ($\Delta\nu$) 
(Hekker et al 2009, Mosser et al. 2010, Bedding et al. 2010, Huber et al. 2010).
Spectra of these stars seem to follow a ``universal pattern'' (Mosser et al. 2011, in agreement with the predictions of the asymptotic  theory of acoustic modes) in  which modes of the same angular degree are distributed along well-defined vertical ridges in the plane $\Delta\nu$ vs. $\nu/\Delta\nu$. The width of these ridges obtained from  CoRoT and {\it Kepler} data show a dependence on the angular degree being the largest for dipole modes in general  
(Bedding et al. 2010, Huber et al. 2010, Mosser et al. 2011b)
and in particular at a value of $\Delta\nu$ corresponding to the luminosity of the Red Clump (RC, see below). As for main sequence solar-like pulsators, it was possible to derive values of the small frequency separations ($\delta\nu_{01}$, $\delta\nu_{02}$)
 (Bedding et al. 2010, Huber et al. 2010) but contrarily to its usual interpretation, their mean values are not related to the properties of the central region (and thus to the stellar age), but to those of the envelope (Montalban et al. 2010a,b).
 The evolutionary state, however can be derived from other features of dipole modes which  are sensitive to the contrast between central and mean densities.  Theoretical predictions  from  stellar models and frequencies, as well as the good agreement between those and observational behaviours,  led to suggest the scatter of dipole ridge in the echelle diagram as diagnostic of the evolutionary state (Montalban et al. 2010a,b).
 Moreover, as  the observation time of red giant increased, the frequency resolution has been noticeably improved and it was possible to  measure the properties of dipole mixed modes (Beck et al. 2011). 
 Following adiabatic and non-adiabatic theoretical  predictions (Montalban et al. 2010a, Dupret et al. 09),
 they should appear around the central peak that corresponds  to the mode mostly trapped in the acoustic (outer) cavity. Bedding et al. (2011) and Mosser et al. (2011a)
  measured the properties of these modes in  {\it Kepler} and CoRoT red-giants as well as their separation in period (what following the asymptotic theory is related to the central stratification), and found that at a given $\Delta\nu$ the value of the period spacing gathers the stars in two different groups. The comparison with  theoretical models allowed their identification as  H-shell burning stars ($\Delta P < 60$~s) and central-He burning ones ($\Delta P > 100$~s).

Important studies have also been done using the basic features of solar-like oscillation spectra : $\nu_{max}$ and $\Delta\nu$. In fact, these quantities are linked  through scaling relations (Brown et al. 1991, Kjeldsen et al. 1995)
 to the values of  global stellar parameters such as mass, radius, and effective temperature.  It is worthwhile remarking that the values of stellar mass and radius resulting from scaling laws and simple features of oscillation spectra are model independent. 
 Miglio et al. (2009) applied these scaling relationships to the  stellar parameter distributions provided by stellar population models to characterize the population of red giants observed by CoRoT. They showed that the population was dominated by low-mass stars in the central-He burning phase (Red Clump stars), and they identified the values  of $\nu_{max}\sim 35\mu $ and $\Delta\nu \sim 4\mu$  at which observational distributions peak (Hekker et al. 2009; Mosser et al. 2010), as those corresponding to low-mass stars at the red clump luminosity. Other studies of populations  have been performed  using the scaling laws, for red giants (Mosser et al. 2010, Kallinger et al. 2010, Hekker et al. 2011)
 and also for the 500 dwarfs and sub-giants pulsators detected by {\it Kepler} (Chaplin et al. 2011).
 The comparison of the mass and radius distributions to the predictions from stellar synthesis populations are very encouraging concerning the validity of these scaling relations and the  perspectives of ``ensemble'' asteroseismology. 

Solar-like oscillations in  RGB (red giant branch)  and RC stars in two stellar clusters have also been studied (Basu et al. 2011).
In particular Stello et al. (2011) 
suggest to use seismic properties as membership diagnostic, and 
Miglio et al. (2011) used these data to test the scaling relations and  estimate mass loss during the luminous RG phase.

Asteroseismology has finally entered  a new era, and because of the unprecedented quality  data, it is now possible to extend the  techniques developed for the Sun to other stars and to perform individual studies of the internal structure. So, the deviations of frequency patterns with respect to the predictions of the asymptotic approximation might indicate the presence of sharps variations in the stellar structure since these features leave a periodic signature in the patter of oscillation modes (Gough 1990). As a striking example, the dependence of $\Delta\nu$ on frequency and the second frequency difference in the oscillation modes of HR~7349 (Carrier et al. 2010)
make evident the presence of a sharp variation in the structure of this red giant. That was identified by Miglio et al. (2010)
 as the signal of the He second-ionization zone observed for the first time in a red giant. The prospect of obtaining precise enough frequencies to derive the helium abundance in the convective envelope of red giants like what  was done in the Sun,  has important obvious   consequences that  overtake  the field  of  asteroseismology.

Seismic studies of individual dwarfs have also provided interesting results, mainly concerning the extension of the convective core and their evolutionary state (i.e., Deheuvels et al. 2010, Deheuvels \& Michel 2010, Metcalfe et al. 2010). 
Moreover, Mazumdar \& Michel (2010) 
detected the signal of HeII in addition to that of the bottom of the convective zone  in the frequency spectrum of HD49933 (Appourchaux et al. 2009, Benomar et al. 2009), 
 an active F-star observed by CoRoT. 

The variability of the characteristics of the p-mode spectrum in the Sun shows a high level correlation with solar surface activity proxies.  This kind of p-mode variations and its correlation with stellar activity has also been observed in  HD49933 (Garcia et al. 2010).
Although the origin of p-mode variability is not completely  understood,  the study of its relationship  with solar activity and with activity cycles of stars with different rotation rates, depth of the convective envelope, and ages will significantly contribute to the understanding of the dynamo processes. 

Finally, solar-like oscillations have also been detected in massive stars (Belkacem et al. 2009a, Degroote et al. 2010) 
such as expected by the theory (Belkacem et al. 2010).
 It is not still clear, however, whether the origin of these oscillations is the convective zone in the iron-peak region or the convective core.

Concerning the self-driven pulsators, significant changes have been observed in the pulsation properties of $\gamma$~Doradus/$\delta$ Scuti stars. While the first ones are classified as pulsating with high order gravity modes and the second with low order acoustic modes and mixed modes, it has been shown that in fact a large fraction of these pulsators shows an hybrid character 
(Hareter et al. 2010, Uytterhoeven et al. 2011).
The interest of these objects is evident since it allows to probe very different regions of the stars in a mass domain where physical processes such as  convection, activity, rotation and chemical peculiarities present rapid changes. The analysis of these pulsators appears at the moment quite complicated given the high number of excited frequencies, Chapellier et al.(2011)  identified 840 intrinsic frequencies in the CoRoT light curve of the $\gamma$~Dor HD~49434.  

  In the same way that the bottom of the convective zone or the second ionization region of helium, the chemical gradient at the border of a receding convective core induces a periodic component in the period spacing of  high-order gravity modes (i.e., Miglio et al. 2008).
  The amplitude and periodicity of this signature, which has been observed in a SPB (slowly pulsating B star) in the CoRoT field (Degroote et al. 2010),
  are directly related to the location and shape of the sharp variation opening the way to the characterization of extra-mixing processes in the central regions of these stars. 

Evolved self-driven pulsators have also taken advantage from the long and continuous time coverage provided by space observations.  First of all, hot sdB are predicted to pulsate with  high order gravity modes (long period ones). These ones are  very useful to probe the deepest regions of the star, but  difficult to observe from ground-based observations.  CoRot and {\it Kepler}  have observed about 15 long period sdB's (Charpinet et al. 2010, {\O}stensen et al. 2010, 2011)
and  the seismic analysis for three of these long period variables suggest that some kind of chemical mixing occurs outside the convective core  (van Grootel et al. 2010a,b; Charpinet et al. 2011).
The  other evolved classical pulsators that have benefited from  continuous time coverage are  RR-Lyrae and in particular the study of the Blazhko effect (Poretti et al. 2010, Chadid et al. 2010, Kolenberg et al. 2011).

With the improvement of observations open questions and new problems arise.  In particular, and as it was already the case
more than 30 years ago, excited pulsations seem to require improvements in the opacity determinations in particular for  the Fe-group opacity peak, where Ni appears to play a relevant role. The difficulties to explain observed frequencies have  driven international collaboration involving theoretical opacity computation teams, laboratory measurements and stellar structure modelers (Turck-Chi{\`e}ze et al. 2011),
as well as  detailed studies of the role of these elements in the microscopic diffusion processes (Hu 2011, Hu et al. 2011).

At the moment,  most of the data  are being interpreted on the base of standard (or quasi) stellar modelling. However, at the same time that this first and fruitful analysis of  CoRoT and {\it Kepler} data is underway, huge effort is being done to test the effect of non-standard processes such as rotation, mixing, magnetic fields, etc.  on the oscillation spectra in order to be able  to go deeper in the analysis and prospects (see \S \ref{transportCC}).

Finally it should be noticed that the connexion seismology-exoplanets is not reduced to the similarity in  the observation requirements. In fact, the capability of seismology in deriving stellar global parameters  is used by the {\it Kepler} team to derive  planet radii, and to provide an estimation of the planetary system age (i.e., Batalha et al. 2011).
An interesting example of the connexion planets-seismology, even if there is no transit, is the multiple system  of sub-stellar objects around the $\gamma$~Dor  HR8799. The identification of the companions as exo-planets or brown dwarfs depends on the age of the system.  As suggested by Moya et al. (2010)
a seismic analysis of  variable host star could provide an estimation of the age, and the space of possible solutions would be reduced if the  inclination of the rotation stellar axis were known. Fortunately, seismology can also provide this information from spectroscopic identification of oscillation modes, such as Wright et al. (2011)
did for  HR8799. 

\section{Low- and intermediate-mass stars (C.Charbonnel)}

We refer to \S \ref{transportCC} for discussions about the impact of non-standard processes in low- and intermediate-mass stars.

\subsection{The Sun (C.Charbonnel)}

During the last decade  the use of 3D solar atmospheric models has lead to a revision downwards of  the abundances of heavy elements (mainly oxygen; Asplund et al. 2005, Grevesse et al. 2010). This has lead to serious discrepancy between available seismic data and 
solar models constructed using these abundances, the main effect of the abundances on solar structure being through reduced opacities in the radiative interior. Considerable effort has been made in order to resolve this discrepancy both by modifying the solar models by changing some of the input physics like diffusion rates (see e.g., Basu \& Antia 2008) as well as by independent determination of abundances.  
One of the solutions that had been proposed to alleviate  the problem with the solar models constructed with AGS05 abundances was to increase the 
abundance of neon since its photospheric abundance is uncertain (Antia \& Basu 2005; Bahcall 
et al. 2005).
Modifications in solar models have not helped in resolving the discrepancy (but see e.g. Turck-Chi\`eze et al. 2010 and references therein). 
Recently, independent 3D models have been used to calculate solar abundances of the key elements and calculated values are higher than the earlier estimates also obtained using 3D atmospheric models (Caffau et al. 2011a).

We would like to emphasize the review by Turck-Chi\`eze \& Couvidat (2011) that illustrates the importance of solar neutrinos in astrophysics, nuclear physics, and particle physics and discusses the many aspects related to the role of rotation and magnetism in solar physics.

\subsection{Mass loss from red giant stars (J. van Loon)}

At the start of 2009, mass loss from red giants posed the following challenges: how can oxygen-rich red giants drive a wind, given the low opacity of their dust? Is dust produced below the tip of the RGB? Does red-giant mass loss increase with metallicity? Some answers have been produced to these questions in the ensuing three years. While a carbon star in the metal-poor Local Group galaxy Sculptor was found to still be rather dusty (Sloan et al. 2009), little more has been revealed to settle the issue whether metal-poor carbon stars produce more or less dust than metal-rich counterparts. Much has been learnt about mass loss from RGB stars in globular clusters, though, with Origlia et al. (2010) finding dust production low on the RGB whereas Boyer et al. (2009, 2010) and McDonald et al. (2009, 2011a,b,c) find dust production to be confined to the tip of the RGB where RGB stars develop cool extended atmospheres and start to pulsate more vigorously. Momany et al. (2011), in an independent study, cast their verdict in favour of the latter. 
McDonald et al. (2010) further offer a new solution to the problem of the missing opacity in oxygen-rich red giant star winds: iron. There remains the question how do the warmer red giants and red supergiants drive a wind. It is very clear now that in metal-poor environments the red supergiants might not become cool enough to drive a dusty wind through combination of pulsation and radiation pressure, and that other mechanisms might need to be invoked such as those operating in chromospherically active stars (Cranmer \& Saar 2011; Mauron \& Josselin 2011; Bonanos et al. 2010) - which are by the way not yet established for certain.

Since the surprising discovery of a long tail of gas trailing the prototypical AGB star, Mira, bowshocks are being found around many red giant stars (Matthews et al. 2011; Mayer et al. 2011; Jorissen et al. 2011; Cox et al. 2011), which is not surprising giving their considerable ages and consequently large peculiar space motions with respect to the interstellar gas. These bowshocks offer a new tool to determine the stellar wind properties of the red giant and to quantify the erosion of the circumstellar envelope (with possible implications for mass-loss rate determinations).

\subsection{Magnetic fields in red giants (C. Charbonnel)}

The red giant branch is a crucial phase for understanding the interplay of magnetic fields and stellar evolution. During the last 3 years, a pilot study of more than 50 red giant stars (mainly selected as presenting evidence of magnetic activity) was carried out with the spectropolarimeters  NARVAL@OHP and ESPaDOnS@CFHT. The first direct detection of the magnetic field of Betelgeuse was made; this magnetic field may be associated to the giant convection cells that could enable a ``local dynamoÓ (Auri\`ere et al. 2010).
Almost all the Zeeman detected sample stars are located at the base of the RGB or at the clump. Importantly the strenght of the magnetic fields was found to be related to the rotational period. The fast rotators host magnetic fields of several to tens Gauss (Konstantinova-Antova et al, 2008, 2009). Magnetic fields as weak as one-half G  were also detected, e.g. in Pollux (Auri\`ere et al. 2009); they correspond to the activity level of solar twins, or even weaker (Petit et al. 2008). For EK Eri, it was concluded that the outstanding magnetic field of about 200 Gauss originates from the remnant of a strongly magnetic Ap star (Auri\`ere et al. 2008, 2011). Konstantinova-Antova et al. (2010) measured the magnetic fields at a level of around a Gauss on several M-type giants among which Ek Boo which is either at the onset of the thermal pulse phase on the asymptotic giant branch, or at the tip of the first red giant branch.  Some of these M stars are known to be rotating fast for their class, and they are presumably also intermediate-mass AGB stars. In this way, a new class of magnetically active stars was unveiled. More observations like these will answer the question whether these stars are a special case, or magnetic activity is, rather, more common among M giants than expected. An unusually high lithium content was reported in HD 232 862, a field giant classified as a G8II star that hosts a magnetic field (L\`ebre et al. 2009).

\subsection{Asteroseismology for red giants (J. van Loon)}
Another field rapidly encroaching upon the study of red giant stars is asteroseismology. First expected to reveal mainly the constitution of main sequence stars like the Sun, perhaps shining light on the age-old problem of convection, it has actually become feasible to use the oscillations of RGB stars to determine their masses (Hekker et al. 2011a,b,c; Miglio et al. 2011). This already opens many ways in which to calibrate stellar models, and one may envisage that in future the detailed oscillation spectrum will become available to study the internal profile of the sound speed much as this was done for the Sun. Much more about Asteroseismology can be found in \S \ref{sect:astero}.

\subsection{AGB and super-AGB stars (J. Lattanzio, J. van Loon, P. Ventura, M. Lugaro, C.Charbonnel)}

Research on AGB stars has continued, becoming more increasingly sophisticated and quantitative but still facing some substantial hurdles, usually associated with convection, convective borders, and other mixing processes. Recent work has significantly advanced our predictions for very detailed neutron capture nucleosynthesis; new models have been published of the s-process in AGB stars both improved and more self-consistent (Church et al 2009, Cristallo et al. 2009, 2011) and for a large range of metallicities (Bisterzo et al. 2010, 2011). These efforts will help us to understand the composition of the elements heavier than iron in Carbon-Enhanced Metal-Poor stars and the production of these elements in low-metallicity environments. 

In connection to the s-process, the first observations of Zr and Rb in putative massive AGB stars (M $>~$ 4 Msun) in the Galaxy (Garcia-Hernandez et al. 2006; Garcia-Hernandez et al. 2007) and in the MC (Garcia-Hernandez et al. 2009) provided the first observational evidence that the Ne22($\alpha$,n)Mg25 reaction is the main neutron source in these stars, as opposed to the C13($\alpha$,n)O16, which is the main neutron source in AGB stars of lower masses. 
 
 Another area that is starting to give up its secrets is the evolution of Super-AGB stars, those which ignite carbon in the core but continue on to experience thermal pulses also (Ventura \& D'Antona 2009, 2010; Siess 2007, 2009, 2010; Doherty et al 2010). The yields from these sources were used to discuss the possibility that these stars played a role in
the self-enrichment of globular clusters, and thus in the formation of multiple populations, as
indicated by recent photometric and spectroscopic evidences (see e.g., D'Ercole et al. 2010 and references therein).

There has always been a lot of activity in the field of Planetary Nebula research, including abundance studies and mineralogy which reflect on the nuclear processing and dust production during the thermal-pulsing part of the AGB evolution (see e.g. Stasinska et al. 2010). The white dwarf mass function has also been a powerful constraint on the AGB evolution. Recently, a new issue has emerged: white dwarf debris discs (Jura et al. 2009a,b; Farihi et al. 2009, 2010a,b; Kilic et al. 2011). Planetary companions are being detected to stars low on the RGB. Where in previous years the focus was on stellar companions and the formation of equatorial density enhancements in the outflows from the AGB progenitor of the PN, possibly due also to shaping by magnetic fields, it has now become necessary to include the effect of planetary systems and debris discs (Farihi et al. 2011).

As the AGB phenomenon spans timescales from a dozen Gyr to as short as dozens of Myr, they are powerful tracers of a galaxy's evolution. Tonini et al. (2009, 2010) and Henriques et al. (2011) have demonstrated that incorporating thermal-pulsing AGB stars in population synthesis codes that are used to model the integrated properties of galaxies makes a notable difference to the interpretation of the evolution of high-redshift galaxies.
Javadi et al. (2011a,b) used the pulsating AGB star population to reconstruct the star formation history of the central part of the Local Group spiral galaxy Triangulum (M33). Applying our knowledge of AGB stars to studies of galaxies is a powerful way to expose and make progress in areas of uncertainty: Javadi et al. found that super-AGB stars must evolve to high luminosities and low effective temperatures in order to yield plausible star formation histories in M33. With extremely large telescopes and the IR-optimised JWST on the horizon, the application of AGB populations to the cosmic evolution awaits a golden age. 

Detailed models of the evolution of isotopic ratios in the Galaxy cannot be complete/correct without including AGB yields (and this point is related to the Vienna conference series). Kobayashi et al. (2011) have produced these predictions with the latest AGB yields and GCE models. 

GCE of isotopic ratios will also help improving our understanding of the evolution of the Galaxy and dust formation when coupled to the composition of stardust grains. Nittler (2009) used the composition of stardust oxide grains to derive a very shallow age-metallicity relationship and to infer that the 18O/17O of the Sun is typical for its age and location (To do this we just need basic
predictions after the first dredge-up) The new models will help making these calculations even more accurate and then, at least in terms of age-metallicity relationships, we can compare the results with stellar surveys (GAIA!). 

\subsection{Binary central stars of Planetary Nebulae (B. Miszalski,
O. De Marco,
A. Acker,
H. Boffin,
R. Corradi,
T. Hillwig, 
D. Jones, 
T. Moffat,
R. Napiwotzki,
J. Nordhaus,
P. Rodriguez-Gil,
M. Santander-Garcia)}

One of the longest standing astrophysical problems concerns finding the mechanism responsible for shaping the extraordinary variety of nebula morphologies (Balick \& Frank 2002). At several recent international meetings researchers in the field have conceded that there is no current theory that can quantitatively predict AGB mass-loss and shaping and hence explain the non-spherical PNe that dominate the PN population. The community therefore rallied around binary central stars as the most promising solution to the problem (for a review, De Marco 2009). During the last three years great strides have been made on both theoretical and observational fronts to determine the extent to which binarity is responsible for the formation and evolution of PNe. 

From an observational perspective, after decades of slow progress we have now more than tripled the number of known close binary central stars to more than 40 post-common envelope systems (Miszalski et al. 2011a); we have independently refined the close binary fraction of Bond (2000) to be at least 17$\pm$5 $\%$ (Miszalski et al. 2009a);  identified strong tendencies for close binaries to be associated with equatorial rings and fast collimated polar outflows (Miszalski et al. 2009b, 2011b; Corradi et al. 2011) and we have identified the binary as the shaping agent in at least 6 systems where the nebula inclination matches the orbital inclination (Jones et al. 2011). In addition several efforts are underway to detect binaries of wider separation to finally determine the total binary fraction of PNe, predicted to be higher than expected from the main-sequence binary fraction (Moe \& De Marco 2006, 2011). 

On the theoretical front, Soker (2006) and Nordhaus \& Blackman (2006) demonstrated that no current theory can explain mass-loss shaping which leads to non-spherical PNe. At the same time, efforts have intensified to demonstrate how common-envelope interactions can shape planetary nebulae (see in particular, Passy et al. 2011). These efforts not only try to determine how binary interactions can shape PNe, but also show that PNe are fundamental tools to constrain common-envelope simulations. This has repercussions for the validity of these simulations in more general contexts, such as the search for a mechanism to explain Type-Ia supernovae.

Our endeavours have blossomed into an exciting and highly active field of research as demonstrated by having, from 2009 onwards, (a) the two highest cited papers in the field (De Marco 2009; Miszalski et al. 2009a), and (b) more than 40 publications. Although we are not yet able to resolve the debate statistically, we would emphasise that binaries can no longer be ignored in the field of PNe and that with the current momentum in the field we are enthusiastically optimistic that future discoveries will bring us much closer to this goal in the medium term.  

Conferences focused on the topic of binarity in planetary nebulae \\
1.Asymmetric Planetary Nebulae 1 (APNI), Oranim, Israel, Aug 8-11, 1994 \\
2.APNII, MIT, Cambridge, USA, Aug 3-6, 1999 \\
3.APNIII, Mount Rainier, Seattle, USA, Jul 28-Aug 3, 2003 \\
4.APNIV, La Palma, Spain, 18-22 Jun, 2007 \\
5.Rochester workshop on Asymmetric Planetary Nebulae, Jun 17-19, 2009 \\
6.APNV, Bowness-on-Windermere, UK, 20-25 Jun, 2010 \\
7.APNVI, Mexico City, Mexico, Summer 2013 (planned)

and their conference summaries: \\
1. Kahn, 1995, AnIPS, 11, 282, Closing summary \\
2.Schwarz, 2000, ASPC, 199, 457, Conference Impression \\
3.Habing, 2004, ASPC, 313, 575, Summary  \\
4.Zijlstra, APNIV conference, 2007, Asymmetric Planetary Nebulae: what are we learning? \\
5.De Marco et al. 2011, APNV conference, 19, The Rochester White Paper: A Roadmap for Understanding Aspherical Planetary Nebulae \\
6.Kastner, APNV conference, 2011, APN V: A Highly Skewed and Biased conference Summary \\

\section{Massive Stars (M.Limongi \& C. Leitherer)}

\subsection{Massive Star Models (M. Limongi \& G. Meynet)}

A good understanding of the evolution of massive stars is required in order to shed light on many topical subjects like
the nucleosynthesis and the UV outputs of the first stellar generations, the origin of the Carbon Enhanced Metal Poor stars, the anticorrelations observed in globular clusters, the properties of the Galactic and the Magellanic Clouds Wolf-Rayet stars,
the final fate of massive stars and how they explode as core collapse supernovae of different types,
the rotation rate of young pulsars and black holes that are produced after the explosion,
the nature of the progenitors of the long Gamma Ray Bursts, the origin of the Be-type stars, and so on. 
In the few lines below, we shall focus on only a few themes on which
substantial works have been made during the 2009-2012 period. 

\subsubsection{Observational highlights}

First, many new observations have been collected to test the theoretical models using technics like
photometry, spectroscopy, interferometry, asteroseismology and spectropolarimetry. 

The use of multiobjects spectrographs has allowed spectrocospic analysis of large populations of massive stars.
Hunter et al. (2009) provide chemical compositions for about 50 Galactic and about 100 SMC early B-type stars in the frame
of the VLT-FLAMES survey. Penny \& Gies (2009) provide velocity values
for 97 Galactic, 55 SMC, and 106 LMC O-B type stars from archival FUSE observations. Huang \& Gies (2010) have obtained
from moderate dispersion spectra the rotation for more than 230 cluster and 370 field B stars. They find that
lower mass B stars are born with a larger proportion of rapid rotators than higher mass B stars.
The spectroscopy of more than 800 massive stars in 30 Doradus (LMC) has been obtained by Evans et al. (2011).  

The Sloan Extension for Galactic Understanding and Exploration (SEGUE) Survey obtained 
$\sim 240,000$ moderate-resolution spectra from 3900 $\rm \AA$ to 9000 $\rm \AA$ of fainter Milky Way stars 
($14.0 < g < 20.3$) of a wide variety of spectral types, both main-sequence and evolved objects, 
with the goal of studying the kinematics and populations of our Galaxy and its halo (Yanny et al. 2009). 
For stars with signal-to-noise ratio $>10$ per resolution element, stellar atmospheric parameters are
estimated, including metallicity, surface gravity, and effective temperature.
The hundreds of stars in this and in the previous Milky Way halo surveys
(HK survey, Beers et al. 1992; Hamburg/ESO (HES) survey, Christlieb et al. 2008)
identified as extremely metal-poor stars (EMP) have been followed up with high resolution spectroscopic 
observations. They provide a means of probing the earliest phases of the evolution of 
the Milky Way and supernovae (SNe) in the early universe.
Extremely important was the discovery of the Galactic halo star SDSS J102915+172927
(Caffau et al. 2011b) with a very low metallicity ($\rm Z\leq 6.9\cdot 10^{-7}$, which is $4.5\cdot 10^{-5}$ times
that of the Sun), $\rm [Fe/H]=-4.99$, and a chemical pattern typical of the classical extremely metal-poor stars - that is,
without enrichment of carbon, nitrogen and oxygen. Such a discovery confirms the previous suggestion,
came with the discovery of the EMP HE 0557-4840 with $\rm [Fe/H]=-4.75$ (Norris et al. 2007), that 
the metallicity distribution function of the Galactic halo does not have a gap between $\rm [Fe/H]=-4.0$, where several stars 
are known, and the two most metal-poor stars, at $\rm [Fe/H]=-5.3$. Moreover, since this star does not show
enhancements of C, N and O, this suggests that the C-richness is not ubiquitous at metallicities below $\rm [Fe/H]=-5.0$, as it was previously though.
In addition to that, the very low global metallicity points toward the crucial role of dust for
the formation of such an extremely metal-poor stars. Caffau et al. (2011b) also claim that
stars similar to SDSS J102915+172927 are probably not very rare and they expect 5$-$50 stars of similar
or even lower metallicity than SDSS J102915+172927 to be found among the candidates accessible from the VLT, 
and many more in the whole SDSS sample. 

Asteroseismology represents an extraordinary window on the interior of stars. Solar-like oscillations
in V1449 Aql, which is a large-amplitude ($\beta$ Cephei) pulsator have been detected by Belkacem et al. (2009a).
Miglio et al. (2009) have used asteroseismic observations to distinguish convective core extension due to overshooting
and the extension due to rotation. Seismic diagnostics of rotation for massive stars have also been studied by Goupil \& Talon (2009).
The stability of g modes in rotating B-type stars has been addressed by Lee \& Saio (2011).  

Interferometry allows to determine the shape of fast rotating stars and to
detect anisotropic distributions of circumstellar matter due for instance to polar winds or to equatorial disk around Be stars. Spectropolarimetry
provides data on the surface magnetic field of early-type stars.
An update survey of these areas of research can be found in the IAU Symposium 279, ``Active OB Stars: Structure, Evolution, Mass-Loss, and
Critical Limits''  (Neiner et al. 2011) and in the recent review
by Walder et al. (2011), ``Magnetic Fields in Massive Stars, Their Winds, and Their Nebulae''.  

\subsubsection{Theoretical developments}
To interpret all these observational data, also theoretical efforts have been made in the last three years in order to
refine as much as possible the theoretical models as well as to include new physical processes like, e.g., stellar
rotation and magnetic fields. 

Massive stars evolve through all the nuclear burning stages until an iron core is formed. This core is surrounded by
the typical onion structure where all the zones keep track of the various nuclear burning occurred either in the core
or in shells. The key aspects of the evolution leading to such a presupernova structure are the following.
1) The neutrino losses due to the pair production become efficient starting from the core C ignition hence the evolutionary 
lifetimes of the advanced burning stages reduce dramatically. 2) All the advanced nuclear burning are activated by few
key reactions that release light particles which can be captured by almost all the nuclei present in the plasma, hence
an increasingly large number of processes become efficient. 3) The nuclear energy generation, which in the core H and core
He burning phases depends essentially on both the temperature and density, becomes very sensitive to the chemical composition
and not only to the temperature and density.
4) Because of the efficient neutrino pair production, the
mixing turnover times become comparable to the evolutionary lifetimes hence a time dependent treatment of convection
becomes necessary. In addition to that, the convective turnover times become of the same order of magnitude of  the
typical nuclear burning lifetimes hence the coupling between convection and nuclear burning become important.
For all these reasons the computation of the presupernova evolution of massive stars requires special attention to the
following points: 1) the equations describing the physical structure of the stars should be coupled to those
describing the chemical composition of the matter due to the nuclear burning and to those for the mixing
due to the various convective processes; 2) a very extended nuclear network, including as much isotopes as possible,
should be adopted and coupled to the above mentioned equations; 3) a time dependent convective algorithm should
be adopted to describe the efficiency of mixing as a function of the time. Different approaches have been
followed by the main groups working on this field in the course of the years. Generally, the equations describing the
physical structure of the star are solved separately to those describing the chemical evolution and
the nuclear energy generation is computed by means of a reduced network (Weaver et al. 1978 and subsequent work)
or in a tabular form (Nomoto \& Hashimoto 1988 and subsequent works). In both cases a quasi- or full- nuclear statistical
equilibrium approximation is adopted above a critical temperature (of the order of $\rm \sim 3\cdot 10^{9}~K$).
Convection is treated as a diffusive process and coupled only to the nuclear burning.
Chieffi \& Limongi were the first to couple both the structure equations to the chemical evolution adopting a 
large network and removing any kind of nuclear statistical approximation (Chieffi, Limongi \& Straniero 1989).
All these computations produced extended set of presupernova models and corresponding explosive chemical yields
that were mainly used for galactic chemical evolution computations as well as supernova light curve and spectra simulations.
The results obtained reproduced the main observational constraints although many open questions remained among which
the reproduction of
(1) the evolution of the abundances of the various elements as a function of the metallicity
in the Milky Way (Francois et al. 2004), (2) the abundance pattern observed in extremely metal-poor stars (Chieffi \& Limongi 2002,
Umeda \& Nomoto 2002, Cayrel et al. 2004), (3) the abundance pattern of the peculiar C-rich ultra metal-poor stars 
(Tominaga et al. 2007). The only way to find a reasonable fit to these observations was to adopt specific fine tuned 
assumptions on both the models and yields.
In the last three years, because of the increasing power of the computer machines, there have been an improvement on
the computational techniques. The american group refined significantly the treatment of the nucleosynthesis, still
separated by the structure equations, by means of an adaptive reaction network including up to $\sim 900$ nuclear
species. Isotopes were added and removed as necessary to follow the nuclear reaction flow, with decisions based upon 
a conservative set of assumptions regarding abundances and flows in the neighborhood (Heger \& Woosley 2010).
Big effort has been done by Chieffi \& Limongi, in the last three years, in order to couple all the equations together 
and solve them simultaneously
in order to avoid any kind of approximation either for the nuclear energy generation and for the coupling
between convection and nuclear burning (Chieffi \& Limongi 2011). 
They adopted a diffusive treatment for the mixing, increased the
size of the nuclear network to $\sim 300$ isotopes and used sophisticated 
state of the art numerical techniques for solving very large sparse systems. This is the best numerical
solution that can be achieved, at present, in 1D stellar evolution codes. Computation of extended grid of presupernova models
for different metallicities are under way (Chieffi \& Limongi 2011).
In spite of these strong efforts made by these groups the ``old" discrepancies between the standard theoretical predictions 
(based on classical models taken at face values) and the observed abundances in extremely metal-poor (normal and C-rich) 
stars still remain. At present no galactic chemical evolution model using these very recent grid of stellar yields
have been carried on hence we have to wait for these results to judge the impact of the improvement in the presupernova models.

The investigation of the potential role of rotation on massive star evolution, started more than 10 years ago 
(Maeder \& Meynet 2000 and subsequent works, Heger, Langer \& Woosley 2000 and subsequent works),
continued in the last three years and is still under way (see also the recent review by Maeder \& Meynet 2011 
and references therein). 

Let us recall that the effects of rotation on massive star evolution can be classified in four categories: 1) axial rotation deforms 
the star and therefore has an impact on the hydrostatic configuration; 2) axial rotation triggers many instabilities in the stellar 
interior driving the transport of the angular momentum and of the chemical species; 3) axial rotation has an impact on the way stars 
are losing mass through radiative winds and through mechanical mass losses; 
4) rotation may in some circumstances activate dynamo mechanisms and thus have an impact on the magnetic field which in its 
turn has an impact on the rotation of stars.  

Through these different effects rotation modifies in a significant way many outputs of the stellar models (evolutionary tracks, lifetimes,
evolutionary scenarios, nucleosynthesis, and so on). 
One of the most important effects for the evolution comes from rotational instabilities inducing efficient mixing of the chemical elements.
Comparisons between the changes of the surface abundances predicted by the rotating stellar models and the observations
are delicate because the observed sample of stars often mix stars of different initial masses and ages. From such comparisons, 
some authors argue that a significant fraction of stars do not follow the trend predicted by the rotating single star models 
(see e.g. Brott et al. 2011). Other
underline that in order for such a relation to be tested great care should be taken to
reduce the range of initial masses and ages in the sample of stars used for the comparison (Maeder et al. 2009).
Probably to make progresses in that field, careful analysis of stars in stellar clusters will be needed.
Stars that  do not follow the expected trends can be due either to close binary evolution
and/or magnetic field (see e.g. de Mink et al. 2009;  Meynet et al. 2011).  

Present stellar models show that all other parameters being kept fixed (initial mass, initial rotation), rotational mixing
is more efficient at low metallicity. This more efficient mixing allows diffusion
of elements between the He-burning core and the H-burning shell at very low metallicity. This boosts the production of some isotopes
like $^{13}$C, $^{14}$N, $^{22}$Ne and the s-process elements and may have an impact on the early phases
of the chemical evolution of galaxies (Cescutti \& Chiappini 2010; Chiappini et al. 2011). It has also been proposed that
the observed ``primary-like'' evolution of Be can be related to the fact that Galactic Cosmic Rays are accelerated from the wind
material of rotating massive stars (enriched by rotational mixing), when the supernova explosion occurs (Prantzos 2010). 
Also fast rotating massive stars have longer main-sequence lifetimes, they
keep a bluer position in the HR diagram and thus may have an impact on the UV outputs. 

The first presupernova rotating models were computed by Heger, Langer \& Woosley (2000).
In this paper they included the centrifugal force following Kippenhahn \& Thomas (1970) in the approximation of Endal \& Sofia (1976)
and the rotational mixing processes discussed in Endal \& Sofia (1978), following the work of Pinsonneault et al. (1989). 
This was an approach different from the one developed by the Geneva group (see above) in the same years.
One of the main results they obtained was that applying the presupernova specific angular momentum of the iron core to a neutron star with
a typical radius of $\rm \sim 10~km$ the newly forming young pulsar would rotate with a period of the order of $\rm \sim 1~ms$,
which is one or even two order of magnitudes larger than the rotational periods of known young pulsars.
Other presupernova models including rotation were computed by Hirschi, Meynet \& Maeder (2004) although their nuclear network
adopted for the advanced burning phases was quite small and included only the alpha isotopes.
Attempts to include additional phenomena in order to slow down the spin up of the iron core were performed by 
Heger, Woosley \& Spruit (2005) which included, in an approximate way, the effect of magnetic breaking.
They found find that magnetic torques decrease the final rotation rate of the collapsing iron core by about a factor 
of 30$-$50 when compared with the nonmagnetic counterparts. However they clearly pointed out that their results
were affected by a number of uncertain parameters. Hence their results could not be considered as well settled.

The study of the interactions between magnetic field and rotation is currently under way and, probably,
will receive even higher attention in the next years. Ud-Doula and Owocki (2009) have 
examined the angular momentum loss of magnetic hot stars with a line-driven stellar wind and a
rotation-aligned dipole magnetic field. These authors give typical spin-down times of the order of 1 Myr
for several known magnetic massive stars. The discovery of rotational braking in the magnetic Helium-strong star
Sigma Orionis E by Townsend et al. (2010) has provided a nice support to the theoretical developments by 
Ud-Doula and Owocki (2009). 
Lau et al. (2011) estimate the spin-down time-scale of rapidly non-convective stars
hosting an $\alpha$-$\Omega$ dynamo. They find that the spin-down time scale
could be only a small fraction of the main-sequence lifetime. 

Very interesting constraints on the question of the magnetic field and rotation will come probably
from complementary observations in asteroseimology, spectropolarimetry and spectroscopy. 
Questions as the following ones could be more precisely addressed: do stars with strong surface magnetic fields
rotate as solid body in their interior? What role does play a surface magnetic field to slow down the central regions?  
What is the role, if any, of magnetic field in the evolution leading to the long gamma ray bursts?
So a very interesting period is ahead of us. 

A big effort was done by Limongi \& Chieffi in order to study the effect of rotation in their more recent set of models
(Chieffi \& Limongi 2011). They included the effect of rotation in their latest most updated version of the
stellar evolution code (FRANEC) following the two proposed schemes, i.e., the one adopted by Heger \& Woosley (2000) and
the one proposed by the Geneva group (see above). A first set of models with and without rotation have been already computed
and will be published very soon. Preliminary results confirm the previous finding of Heger \& Woosley (2000) that
these models have even much more angular momentum in the collapsing iron core than a neutron star can possibly carry, i.e.,
too fast rotating pulsars will be produced by these models. This means that still a lot of work must be done in order
to revisit the mechanisms of the angular momentum transport (including the interaction between rotation and magnetic fields)
during the presupernova evolution of massive stars.

\subsection{Massive Binary Evolution (S. de Mink)}

In the last three years there has been an increasing interest in the importance of binarity for the evolution of massive stars. 
This is partially driven by new determinations of the high close binary fraction in massive stars, such as performed by Sana \& Evans (2011).  
These authors compiled the most complete dataset of the binary properties of young massive stars in nearby clusters.  
A striking result is the strong preference of for binaries with orbital periods of several days and less. The authors convincingly 
argue that this cannot be attributed to observational biases alone, as was previously thought.  
Questions about the role of binaries where also raised in the context of the puzzling trends of observed rotation rates and 
surface abundances found in the VLT-FLAMES survey of massive stars (Hunter et al. 2008; Evans et al. 2011), even for stars that 
appear to be single (de Mink et al. 2011). 

Studies of individual massive binary systems are now providing us with ever more accurate parameters of the most massive stars, 
enabling critical tests for evolutionary models in the upper parts of the HR diagram (Pavlovski \& Southworth 2009; Pavlovski et al. 2009; 
Ritchie et al. 2009; Ritchie et al. 2010; Clark 2011).  Record holders include R136,  host of the most massive O supergiants 
(Taylor et al. 2011), and R145 (Schnurr et al. 2009) hosting a hydrogen-rich Wolf-Rayet star for which minimum masses near $\rm 100~M_\odot$ 
have been derived.   

Significant progress has also been made with respect to modeling the evolution of stars in binaries. 
Cantiello et al. (2007) and de Mink et al. (2009) explored the consequences of rotationally induced mixing on stars in binaries. 
Eldridge et al. (2008),  Yoon et al. (2010) and Claeys et al. (2011) focussed on the consequences of binary interaction on the progenitors 
of supernova.  Eldridge et al. (2011) undertook an ambitious population synthesis study of runaway stars as the progenitors of 
supernova and gamma-ray bursts.

\subsection{Existence of Extremely Massive Stars in R136 (P. Crowther)}

Until recently, a general consensus had been reached that the upper stellar mass limit was close to $\rm 150~M_\odot$ (Figer 2005). 
Crowther et al. (2010) reopened this debate, by attributing initial masses of up to $\rm \sim 300~M_\odot$ for the brightest stars 
within R136, the central ionizing cluster of the 30 Doradus star forming region in the LMC. Crowther et al. attributed higher 
stellar masses than previous estimates for these stars using new infrared VLT spectroscopic and photometric observations, 
together with high stellar temperatures from spectral analyses and contemporary evolutionary models for very massive stars. 
Supporting evidence for their results was provided by consistent spectroscopic and dynamical mass estimates for NGC~3603-A1, 
a massive eclipsing binary system located in a similar star cluster within the Milky Way (Schnurr et al. 2008).

If the stellar mass limit were to exceed $\sim 150 M_\odot$, this would open up the possibility of pair-instability supernova -- 
hitherto thought to be restricted to Population~III stars -- within the local universe. Indeed, Gal-Yam et al. (2009) have 
attributed SN 2007bi, an extremely bright supernova,to the pair-instability supernova of a metal-poor $\sim 200 M_\odot$ star. 
The high luminosities and powerful stellar winds of very massive stars would be anticipated to dominate the early appearance and 
feedback from high-mass star clusters. Indeed, Hoversten \& Glazebrook (2011) argue that stars more massive than 120 $M_\odot$ are 
required to reproduce the observed H$\alpha$ to continuum ratio of Sloan Digital Sky Survey galaxies.

\subsection{Luminous Blue Variable-type Eruptions (N. Smith)}

Luminous Blue Variable (LBV)-like eruptions have been recognized as playing a more dominant mass-loss role than previously 
thought for two reasons.  One reason is that the inclusion of far-IR observations (sampling cool dust) has shown that LBV shells 
contain far more mass that previously known (e.g., Smith \& Owocki 2006).  In some extreme cases, an LBV can shed as much as 
15-20 $M_\odot$ in a single eruption lasting only a few years, and these can happen multiple times in a star's life. A large number 
of new dust shells around LBVs and related stars have been discovered in recent surveys with Spitzer, showing that the phenomenon 
is widespread (Wachter et al. 2010; Gvaramadze et al. 2011).  The second reason is that a large amount of work on the steady 
line-driven winds of hot O-type and Wolf-Rayet stars has revealed that their winds are very clumpy, and that standard mass-loss 
rates that have been used for decades are therefore too high by factors of perhaps 3 -- 10 (Bouret et al. 2005; 
Fullerton et al.\ 2006; Puls et al. 2006). This indicates that while steady, metallicity-dependent winds are less important in 
removing a massive star's H envelope, LBV eruptions (which may be independent of metallicity) are probably the dominant mode of 
mass loss (Smith \& Owocki 2006).

Studies of a particular type of supernova called Type IIn supernovae (the ``n'' is for narrow H lines) reveal that eruptive 
LBVs might be the progenitors of some of the most luminous supernovae known (Smith et al.\ 2007).  This is a surprise, since 
LBVs are generally expected to shed their H envelope and evolve to the Wolf-Rayet phase before exploding (Heger et al. 2003).  
These Type IIn supernovae represent about 8 -- 9\% of all core-collapse supernovae (Smith et al. 2011a), but they may be the 
most massive 8 -- 9\% of stars that explode.  The key observation is that the circumstellar material into which the supernova 
blast wave expands is extremely dense and massive, requiring eruptions of large amounts of mass just a few years or decades prior 
to core-collapse (Smith et al. 2007, 2008, 2010; Ofek et al.\ 2007; Woosley et al.\ 2007; Chevalier \& Irwin 2011).  The only 
known analog to produce the required amount of mass loss in such a short time is the eruptions of LBVs.  There have been three 
cases where a IIn progenitor star candidate was identified in pre-explosion archival images, and those sources are consistent with 
very massive blue LBV-like stars (Gal-Yam \& Leonard 2009; Smith et al. 2011b; Smith et al. 2011c). Therefore, it would seem 
that LBV-like eruptions can be an immediate precursor to a core-collapse supernova for very massive stars. 

Recent years have seen the discovery of a substantial number of extragalactic eruptions that resemble LBV eruptions 
(i.e. fainter than supernovae and similar spectra), partly due to the increased emphasis on transient studies and the 
different methods employed in supernova surveys.  These have yielded quite surprising results.  One of the most interesting 
is that the family of eruptive outbursts that has been referred to as LBVs is surprisingly diverse, and some of them appear 
to come from rare classes of progenitor stars that are not necessarily the most massive stars (Thompson et al. 2009).  
This diverse class of transients is referred to variously as LBV eruptions, $\eta$~Car analogs, supernova impostors, 
intermediate-luminosity optical transients, and luminous red novae (see Smith et al. 2011b and references therein for a recent summary).  
The underlying physical mechanism remains unknown, and various possibilities have been proposed, including super-Eddington 
LBV eruptions, binary merger outbursts or stellar collisions, electron capture supernovae, nuclear flashes, and ``failed'' 
supernovae (underluminous supernova from fallback to a black hole).  Some of the progenitor stars of these outbursts certainly 
are what we would call classical LBVs, but some appear to be heavily dust enshrouded red stars with luminosities that suggest 
initial masses as low as 8~$M_\odot$ (Thompson et al. 2009; Prieto et al. 2009).  Yet, the observed outburst properties are all 
quite similar and resemble known LBVs.

\subsection{CNO Anomalies in Massive Stars (D. Lennon)}

Our understanding of the origin of CNO anomalies in main-sequence massive stars has undergone what might be described as a 
paradigm shift in recent years. This was prompted in part by the VLT-FLAMES Survey of Massive Stars which found little evidence 
for a correlation between nitrogen  enrichment and stellar rotation (Hunter et al. 2009).  Their work presented surface abundances 
of fast rotators for the first time. However the most extreme cases of nitrogen enrichment were found among the slowest rotators, 
challenging our ideas concerning rotational mixing and leading to new estimates of the efficiency of rotational mixing and convective 
overshooting (Brott et al. 2011). The picture is further complicated  by the realization that some nitrogen enriched slowly rotating 
B-type main-sequence stars in the Galaxy are also magnetic (Morel 2009) although the link between magnetism and nitrogen enrichment 
is as yet unclear. What is now evident however is that for those OB stars with CNO anomalies we are indeed seeing the products of  
CNO-cycled matter mixed to their surfaces, as shown by Przybilla  et al. (2010) in a very careful analysis of Galactic B-type stars. 
A quantitative understanding of CNO anomalies in more massive  early O-type stars has been hindered by limitations in non-LTE models 
of the line formation of higher ionization stages of nitrogen in particular. It is therefore gratifying to see crucial progress being 
made on this front, Rivero et al. (2011) presenting a comprehensive analysis of the N\,{\sc iii} line formation problem. 

Hunter et al. (2009) also found that among their fast rotators there exists a wide range of nitrogen abundances, a result which 
was reflected in related work by Dunstall et al. (2011) for Be stars. The lack of a strong correlation between rotational velocity 
and nitrogen enhancements in massive stars, together with almost bimodal distribution of rotational velocities, has led to speculation that
binaries may play a more important role than was previously thought in the evolution of massive star populations 
(de Mink et al. 2009a, even to the extent of perhaps explaining abundance anomalies in globular clusters (de Mink et al. 2009b). 
As discussed above, Sana \& Evans (2011) emphasize the importance of the binary channel in massive star evolution, showing that 
approximately half OB stars in open clusters are in fact spectroscopic binaries. Additionally, it is suggested that binary 
interaction may produce a system which is difficult to distinguish from a single star (de Mink et al. 2011) which implies that  
understanding the origin of CNO peculiarities in an apparently single star may hinge on distinguishing between single and binary 
star evolutionary histories. Complicating the picture even further, Pavlovski et al. (2011) analysed  disentangled spectra of 
components of high-mass binaries and have so far found no trace of CNO cycled material on their surfaces. 

In summary, the processes leading to the origin of CNO anomalies  in OB stars are much more complex than thought only a few years ago. 
Ongoing large scale observing programs such as the VLT-FLAMES Tarantula Survey (Evans et al. 2011), the VLT-FLAMES survey of 
massive binaries in Westerlund I (Ritchie et al. 2009), and MiMeS Project (Wade et al. 2011) surveying for magnetic fields in massive stars (see below) offer the potential to shed light on this problem.

\subsection{Magnetism in Massive Stars (S. Owocki \& G. Wade)}

Hot, massive OB stars lack the vigorous subsurface convection thought to drive the dynamo central to the magnetic 
activity cycles of the sun and other cool stars. But building on pioneering detections of strong (kG) fields in the chemically 
peculiar Ap and Bp stars (Babcock 1947; Borra \& Landstreet 1980), new generations of spectropolarimeters have revealed 
organized (often significantly dipolar) magnetic fields ranging in strength from 0.1 to 10 kG in several dozen OB stars 
(e.g. Donati et al. 2002, 2006; Hubrig et al. 2006; Grunhut et al. 2009; Martins et al. 2010; Petit et al. 2011). 
A consortium known as MiMeS (for Magnetism in Massive Stars) is carrying out both a survey for new detections, and monitoring 
known magnetic OB stars with high resolution spectroscopy and polarimetry (Wade \&  the MiMeS Collaboration 2010). 
The observed field characteristics favor a fossil origin over active dynamo generation (e.g., Donati \& Landstreet 2009). 

Much current research focuses on the effects of the magnetic field on stellar mass loss, rotation, and evolution. 
Theoretical models and MHD simulations (e.g. Townsend \& Owocki 2005; Ud-Doula et al. 2008, 2009) indicate magnetic trapping 
of stellar wind outflow can feed a circumstellar magnetosphere, resulting in rotationally modulated Balmer line emission 
that matches closely that which is observed (e.g., Townsend et al. 2005; Oksala et al. 2011). Moreover, eclipses by 
co-rotating magnetically bound clouds induce a photometric variation that, with extended monitoring, allows precise 
measurement of the stellar rotation, and even its secular slowing due to angular momentum loss from the magnetized wind. 
For the B2V magnetic star $\sigma$~Ori~E, the inferred spindown time of 1.3~Myr is in very good agreement with theoretical 
predictions based on MHD models (Townsend et al. 2010). And the discovery of very long rotation periods (years to decades) 
in two O-type magnetic stars (Howarth et al. 2007; Naz\'e et al. 2001) provides further evidence that magnetic braking can 
strongly affect the rotational evolution of magnetic massive stars. Larger population samples are needed to understand the 
origin of these strong stellar fields, to determine their long-term evolution, and to clarify their potential role for 
massive-star evolutionary end-states as SNe, pulsars, and magnetars. 

\section{Supernovae (M. Limongi, A. Tornamb$\rm \bf \grave{e}$, A. Mezzacappa)}
A substantial fraction of the huge number of papers published on supernovae during the past three years, i.e. more than 5000,
are related to the two main still open questions: (1) the nature of the progenitor of the thermonuclear supernovae (SN Ia) and
(2) the comprehension of the explosion mechanism for core collapse supernovae.

The nature of the progenitors of type Ia supernovae still remains elusive
in spite of the huge efforts to identify the(ir) evolutionary path(s) produced  
in the last three years by several authors. Needless to recall
the pivotal role of type Ia SNe in cosmology and therefore the need to clarify
in detail all their evolutionary and explosive properties.
The historical progenitor models are essentially two: 
the single degenerate (SD) model, consisting of a normal star accreting H on a CO white dwarf (WD) 
(see  Kobayashi \& Nomoto 2009 for an updated version of this model) 
and the double degenerate model (DD), consisting of two WDs which merge due to 
gravitational radiation emission. Both SD and DD models suffer drawbacks.
One problem for the DD scenario largely addressed in the last years arises
when the two dwarfs strongly interact in a very short time during the merging.
A number of authors have followed the dynamical evolution of such a system in order to identify the outcome and to asses 
how much the classical view of the secondary dwarf dissolved in a disk, which provides mass to an 
almost untouched primary, has to be modified  (Loren-Aguilar, Isern and Garcia-Berro 2009, Pakmor et al. 2011). 
This aspect deserves however further attention.
Stellar rotation has been claimed to be the pivotal mechanism 
leading to a self-regulated accretion rate during the
merging of the two components which, in turn, tightly points toward an explosive outcome.
All along this evolutionary path an additional source of gravitational waves emission has been
also found to exist. These studies were carried on in the assumption of the solid body
rotation (Tornamb$\rm \grave{e}$ et al 2004). Extension to differentially rotating
accreting WDs suggests to revisit the evolutionary path toward the explosion.
Such a differentially rotating structure may still remain stable 
even if the mass largely overcomes the Chandrasekhar limit
(which is $\rm \sim 1.5~M_\odot$ for a solid body rotation). 
Therefore, supra-stable structures
may be produced during the accretion phase. These structures will become suddenly unstable when the
accretion phase ends and the angular momentum is redistributed all along the accreted structure.
In this scenario there is a change of perspective in the description of the evolutionary path 
toward the degenerate explosion: 
we are no more facing a degenerate structure which is driven to the critical limit by mass accretion but,
on the contrary, a smooth process in which the critical mass is safely and largely bypassed 
until the structure become suddenly unstable when the accretion phase ends (Piersanti et al. 2009). 
Pfannes et al. (2010a,b) mimicked the explosion of a fast rotating massive WD obtaining results
which, in the case of C-detonation, provide a good agreement with the observations.
Maeda et al. (2010) explain the diversity of the spectral evolution of type Ia SN in the framework of an 
asymmetric explosion which is the natural outcome of the explosion of a fast rotating structure.     
In the case of a SD progenitor, rotation has also been invoked to suppress
the strong nuclear pulses occurring when the accreted hydrogen
is converted into He and then He into C that, in turn, would avoid the growth to the Chandrasekhar 
limiting mass because of strong mass loss occurring during the pulses. 
All in all, while rotation helps to trace the good route to the explosion, 
no robust effect to discriminate theoretically DD from SD models
as progenitors of SNe Ia has been found yet.
Interesting observational results seem however to point toward progenitors mostly composed by DD systems
(Gifanoy \& Bodgan 2010). The reason is that the observed X-ray flux from nearby elliptical galaxies is a factor of $\sim 30-50$ 
less than the one predicted in the case of the H-accretion SD scenario.
Several observational approaches have been suggested to discriminate the nature of the progenitor model 
(among them, Fryer et al. 2010, Bianco et al. 2011, Livio \& Pringle 2011, Marsh 2011) which should be
exploited in the near future. At present, the diversity of the type Ia SNe has been emphasized in some details by means of
HST observations in a GO international project whose results will soon appear on ApJ (Xiaofeng Wang et al. 2011).
Once again non-asymmetric explosion are claimed to explain observations. 
A very interesting observational result comes from the observations of
the near-by SN Ia in M101 (2011fe). According to Li et al. (2011) the hypothesis that a red giant star 
could have been present in the progenitor system has to be ruled out from the analysis of the archive frames 
obtained prior to  the explosion occurred in August 2011. The progenitor system was therefore
made of two WDs, i.e. consistent with the DD scenario, 
or, at most, by a WD plus a Main Sequence (or very little evolved) star companion.
None of the described results (neither observational nor theoretical) seem to be able, however, to challenge
the cosmological use of the SN Ia, as it was done so far. 

The past several years have seen notable progress toward ascertaining the core collapse supernova explosion mechanism. 
Three independent groups worldwide (Bruenn et al. 2009, Marek \& Janka 2009, and Suwa et al. 2010) have now reported 
neutrino-driven explosions, beginning with a range of stellar progenitor masses from 11 to 25 $\rm M_\odot$. 
In all cases, the standing accretion shock instability (SASI; Blondin, Mezzacappa, \& DeMarino 2003) couples 
with neutrino heating to power the explosions and thus plays a central role. These results were obtained in 
two-dimensional simulations. Key now will be an extension of these models to three dimensions. 
All of the above are multi-physics models. In particular, they all deploy multi-frequency neutrino transport. 
In addition, approximate general relativity is included in some of the models, as well as state of the art neutrino 
weak interactions and industry-standard nuclear equations of state. While WilsonÕs delayed-shock mechanism (Bethe \& Wilson 1985) 
remains operative, significant variations on this theme have been reported in the above models. In particular, the explosions 
develop over a significantly longer period of time than originally reported by Wilson, with weak explosion energies at 
times significantly less than one second after stellar core bounce and still growing as late as one second after bounce. 
The SASI leads to continued long-term accretion to deep layers above the proto-neutron star surface, which in turn leads 
to efficient neutrino heating of the accreted material, some of which becomes unbound and joins the ejecta. The neutrino 
transport approximations used in the above models Ð in particular, the Òray-by-rayÓ approximation (Buras et al. 2003) Ð must be 
replaced by two- and three-dimensional multi-frequency, and ultimately multi-angle and multi-frequency, neutrino transport. 
In three spatial dimensions, multi-angle and multi-frequency neutrino transport will require sustained exascale supercomputing 
resources to complete. The approximation to general relativity (Marek et al. 2006) used in the above models must also be replaced, 
especially if peculiar core collapse supernovae are to be studied. The ability to evolve black-hole-forming collapse 
will be essential in order to understand systems with significant rotation and magnetic field generation and their connection 
with hypernovae and gamma-ray bursts. With non-parameterized explosions now in hand, the opportunity presents itself to compute 
the gravitational wave signatures from core collapse supernovae, through explosion (Yakunin et al. 2010). While these computations 
are based on two- and not three-dimensional models, they mark the first time all phases of core collapse supernova gravitational 
wave emission have been delineated in non-parameterized models, providing more accurate prediction of the development of the gravitational 
wave amplitudes and more accurate prediction of the time scales over which the amplitudes develop. These computations must now 
be repeated once comparable three-dimensional models are in hand. 

\section{Stellar populations (E. Brocato)}
Stars are the basic elements of the light emitted by stellar clusters and galaxies, so that the challenge of de-convolving the  
light emitted by such distant objects have been caught by stellar population synthesis models which, typically, foresee the expected  
integrated photometric colors and spectra.
An innovative method to use the light emitted by stars to unveil information on the age and metallicity of stellar populations 
in unresolved galaxies is the multi-band measurements of their Surface Brightness Fluctuations (SBF). 
As suggested by stellar population synthesis models (Worthey 1993; Brocato et al. 1998),  
the last years have been made clear that SBF are an independent and effective tool to derive information on the features of 
stellar populations, in particular when optical and near-IR high-quality images are observed (e.g. Blakeslee 2009; 
Cantiello et al. 2011). An exiting new branch of this technique discloses the possibility of deriving the age of 
stellar clusters both on theoretical and observational bases (Raimondo 2009;  Whitmore et al. 2011).

\section{Closing remarks}

We wish to thank the researchers who actively contributed to this report, and apologize to those whose work is not quoted here. Progress has been made in many other fields within our remit that unfortunately we cannot all do justice here. As a matter of fact, C35 researchers are extremely active and their work has tremendous impact on all the areas of astrophysics and cosmology. It is thus an impossible task to present an exhaustive review of all the results.

\vspace{3mm}
 
{\hfill Corinne Charbonnel}

{\hfill {\it president of the Commission 35}}

\end{document}